# Recruiting Hay to Find Needles: Recursive Incentives and Innovation in Social Networks

Erik P. Duhaime[1], Brittany M. Bond[1], Qi Yang[1], Patrick de Boer[2], & Thomas W. Malone[1]

[1]Massachusetts Institute of Technology
[2]University of Zurich

Finding innovative solutions to complex problems is often about finding people who have access to novel information and alternative viewpoints. Research has found that most people are connected to each other through just a few degrees of separation, but successful social search is often difficult because it depends on people using their weak ties to make connections to distant social networks. Recursive incentive schemes have shown promise for social search by motivating people to use their weak ties to find distant targets, such as specific people or even weather balloons placed at undisclosed locations. Here, we report on a case study of a similar recursive incentive scheme for finding innovative ideas. Specifically, we implemented a competition to reward individual(s) who helped refer Grand Prize winner(s) in MIT's Climate CoLab, an open innovation platform for addressing global climate change. Using data on over 78,000 CoLab members and over 36,000 people from over 100 countries who engaged with the referral contest, we find that people who are referred using this method are more likely than others both to submit proposals and to submit high quality proposals. Furthermore, we find suggestive evidence that among the contributors referred via the contest, those who had more than one degree of separation from a pre-existing CoLab member were more likely to submit high quality proposals. Thus, the results from this case study are consistent with the theory that people from distant networks are more likely to provide innovative solutions to complex problems. More broadly, the results suggest that rewarding indirect intermediaries in addition to final finders may promote effective social network recruitment.

RECURSIVE INCENTIVES AND INNOVATION IN SOCIAL NETWORKS

# 1 INTRODUCTION

## 1.1 Open Innovation

Advances in communication technologies have made it easier than ever to harness the collective intelligence of large groups of people through crowdsourcing. For instance, crowdsourcing can be used to collect product opinions or valuable information about recent events and also to mobilize strangers toward collective action [1-3].

Open innovation contests—a particularly impactful form of crowdsourcing—are designed to attract creative solutions to difficult problems from a large number of potential solvers [4]. By opening up a problem to thousands, millions, or eventually billions of people around the world, problem broadcasters can reap the benefits of vast distributed knowledge and also increase the chances that they will be able to find a "needle in a haystack" solution to a complex problem [5].

Importantly, research on open innovation contests has found that such contests work not only because problem broadcasters are provided with potential solutions from *more* people, but also because they receive many ideas from very *different* people. For instance, in analyzing a dataset on 12,000 scientists' activity in 166 contests on the open-innovation site Innocentive (www.Innocentive.com), Jeppesen and Lakhani (2010) found that the best ideas came both from people who are often socially marginalized, specifically women, and from people whose technical expertise was in a field different from the problem at hand (e.g., a chemist solving a biology problem) [6]. Relatedly, Duhaime, Olson, & Malone (2015) found that ideas submitted to MIT's Climate CoLab (www.climatecolab.org) —a website where anyone in the world can submit solutions to climate-related issues—were just as likely to come from women as from men, from residents of other countries as from the US, and from people without a graduate school education or experience with climate issues [7].

Since the advantages of open innovation contests over traditional means of innovating are driven by attracting not just more people, but more people with access to novel information and perspectives, a central question then becomes: what is the best way to recruit the best solvers?

## 1.2 Social Network Recruitment

Many organizations rely on social network recruitment for finding key individuals. For instance, employee referral programs are a common way for employers to tap into the potential value of current workers' social networks, and recruitment by word-of mouth is the most common method organizations use to attract job applicants [8-10]. This literature has illuminated how "weak ties" (i.e., relationships to distant acquaintances rather than close associates) often enable access to novel information because they are more likely to bridge "structural holes" between social networks [11-12]. Applying this insight to the domain of finding solvers for open innovation contests, we reasoned that activating weak ties would be especially important for recruiting the best solvers to an open innovation contest.

## 1.3. Motivating Weak Ties with Recursive Incentives

While weak ties may be the most important for successful social network recruitment, they can also be very difficult to motivate [13-14]. Furthermore, social network search often requires exploring several different paths of social links, and motivating weak ties may become increasingly difficult with increasing social distance [15]. For instance, research on social search (i.e., finding specific distant individuals through successive network tie activation) has also highlighted the importance of weak ties and has shown that target individuals can be "found" several degrees of separation away, but such success is highly sensitive to the individual incentives of distant weak ties [16-17].

One promising approach for motivating weak ties to boost social network recruitment was demonstrated by Pickard et al (2011), who won the 2009 DARPA Network Challenge in which teams competed to be the first to identify the locations of ten red weather balloons placed at ten different previously undisclosed locations around the United States [18]. The team implemented a "recursive incentive mechanism" where people were rewarded not only for identifying a balloon, but also for referring others who helped to identify a balloon.

The approach, depicted in Figure 1, had several desirable attributes that motivated people to spread the word about the contest. First, it reduced disincentives stemming from competition (i.e., the incentive to not tell others, lest they find the balloons instead). Second, it broadened the scope of people that one might refer to the contest. Whereas a traditional referral incentive would motivate people only to spread the word to people who they think might be directly interested, the recursive incentive scheme provided an incentive to share the contest with anyone who might be indirectly interested (i.e., interested in sharing the contest with others). In this way, the recursive incentive scheme incentivized people to spread word of the contest more broadly and also motivated intermediaries to continue spreading the word, enabling information about the contest to reach distant networks.

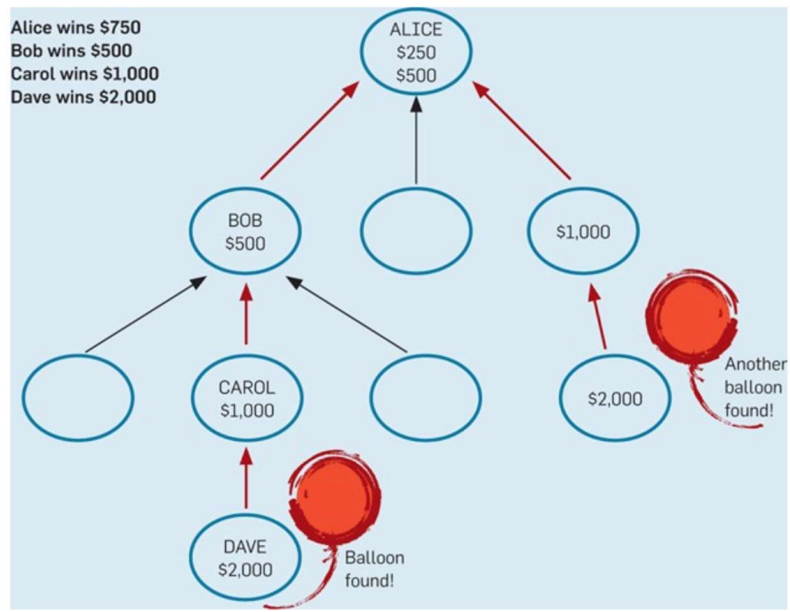

Figure 1: A figure used by the MIT Red Balloon Challenge team to explain the recursive incentive structure. Alice Joins the team and is given an invite link, like http://balloon.mit.edu/alice. Alice then emails her link to Bob, who uses it to join the team as well. Bob gets a unique link, like http://balloon.mit.edu/bob, and posts it on Facebook. His friend Carol sees it, signs up, then tweets about http://balloon.mit.edu/carol. Dave uses Carol's link to join, then spots one of the DARPA balloons. Dave is the first person to report the balloon's location to the MIT team, helping it win the Challenge. Once that happens, the team sends Dave $2,000 for finding the balloon. Carol gets $1,000 for inviting Dave, Bob gets $500 for inviting Carol, and Alice gets $250 for inviting Bob.

Here, rather than use a recursive incentive scheme to motivate people to find balloons, we used it to motivate people to find people with the best ideas for an open innovation contest. We reasoned that, like with the Red Balloon Challenge Team's approach, a recursive referral incentive would motivate people to spread word of the contest to weaker ties (i.e., not only people who might have an idea, but also to people who might know someone with an idea), thereby spreading word of the open innovation contest to more distant, heterogeneous social networks. In addition to bringing in more recruits, we hypothesized that many of these recruits would bring novel information and perspectives and as a result submit innovative ideas.

## 2 RESEARCH SETTING

### 2.1 MIT Climate CoLab

The MIT Climate CoLab (www.climatecolab.org) software platform allows individuals and teams of people from anywhere in the world to develop proposals for how to address global climate change [19-23]. As of July 2017, when data collection for this study ended, over 600,000 people from virtually every country in the world had visited the CoLab site, over 78,000 had registered as members, and over 1,250 had contributed to at least one proposal. (Today, there are over 100,000 members.)

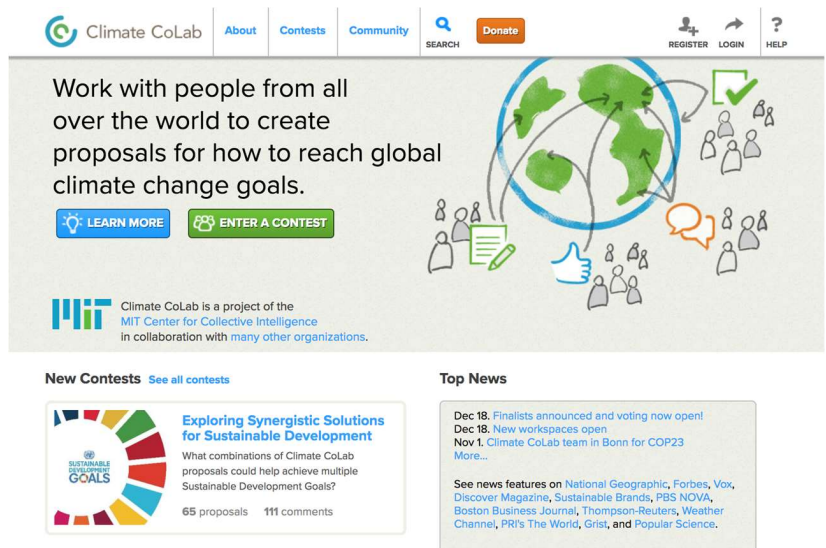

Figure 2: The MIT Climate CoLab homepage

Activity on the CoLab platform is driven by a series of annual contests (typically over a dozen) on topics such as how to reduce emissions in the transportation sector and how to change public attitudes about climate change. Some of the contests are run in conjunction with organizations such as the Union of Concerned Scientists (www.ucsusa.org) and the Carbon War Room (www.carbonwarroom.com). Each contest has advisors and judges, including experts from organizations like NASA, the World Bank, MIT, and Stanford, as well as one former US Secretary of State (Shultz), two former US Congress members (Inglis and Sharp), and two former heads of state (Robinson and Bruntland).

After proposals are submitted, the judges select the most promising entries to be semi-finalists, provide feedback to help improve the semifinalist proposals, and later select finalists. Then from the finalist proposals, the judges select the Judges' Choice Awards and the community votes for the Popular Choice Awards. The Climate CoLab offers a cash award—typically $10,000—to one Grand Prize Winner. All the Popular and Judges' Choice winners receive opportunities to present their ideas to top experts and potential implementers, usually at the Crowds & Climate conference, which was held at MIT annually from 2013-2016.

## 2.2 The Social Network Prize

Beginning with CoLab's annual contest of April 2014, we launched a referral contest with recursive incentives that we dubbed the "Social Network Prize." Each time someone clicked on a link for the competition they went to a landing page (Figure 3, below) explaining the contest. If a user clicked on the blue hyperlink to "see how it works," they were provided with an example based on the MIT Red Balloon Challenge Team's example presented in Figure 1. Specifically, it read:

It might play out like this: Alice enters her e-mail address below and we give her a unique invite link. Alice then e-mails her link to Bob, who also enters his e-mail address below so that we can give him a unique invite link, which he then posts to Facebook. Bob's friend Carol sees the link, signs up to get her unique link, which she then posts to Twitter. Carol's follower Dave then sees the link, signs up for the CoLab and creates (or helps to create) the best-ranked proposal, winning him the Grand Prize of $10,000. Once that happens we send Carol $1,000 for inviting Dave, Bob gets $500 for inviting Carol, and Alice gets $250 for inviting Bob. If the chain of friends were even longer, then we would give out another $125, $67.50, and so on.

Upon reaching the landing page, if a person wanted to explore the CoLab site and become a member, they could easily do so. Even if they were not interested in becoming a member, they could still refer other people to the CoLab site and thus become eligible to win some of the Social Network Prize. To accomplish this we needed their email address so that we would be able to give them their reward should they win it. Entering their email address generated a unique link associated with their ID number that they could send to people through twitter, Facebook, e-mail, or by any other means.

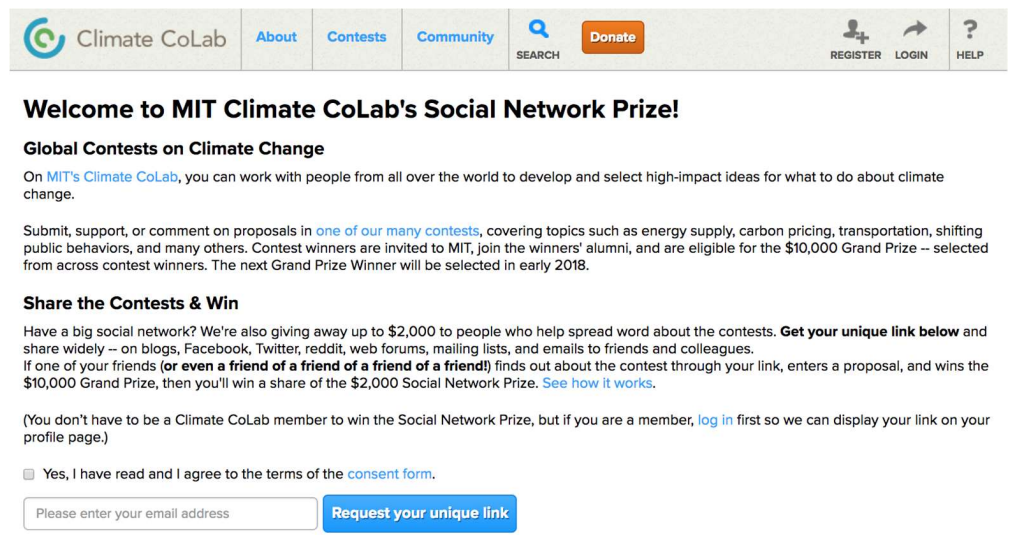

Figure 3: The landing page of the Social Network Prize contest. After agreeing to the consent form and entering their email address, people would be provided with a unique link that they could share with others. Once someone clicked on that link, they would arrive at this same landing page.

From an initial starter link for the competition as a whole, we created unique links for staff members to post on the CoLab website and to share in messages to all CoLab members, in blog posts, and in posts to various social media forums like Twitter, Facebook, and Reddit. Notably, staff members were not allowed to win the contest and therefore did not have a personal financial incentive to share with their networks. Thus, links were largely disseminated through the same channels that staff members typically used for other information about the CoLab. Staff members occasionally created and shared new links in order to disseminate them in different places, which would artificially inflate the size and complexity of the network. Therefore all links generated by a staff member were condensed into a unique staff referral link for analysis purposes.

Identifying unique chains of referral links was straightforward since everyone had to click on a unique link from someone else and then generate a unique link in order to share with others. However, we also needed to be able to identify whether a new Climate CoLab member was a Social Network Prize recruit (even if they did not themselves enter their email address to create their own unique link, which might happen if the new member was interested in the CoLab but not in recruiting new participants). Relatedly, when someone clicked on a Social Network Prize link, we needed to identify whether they were a new recruit or an existing member. We accomplished this by using browser cookies. Specifically, we installed a browser cookie when someone landed on the Social Network Prize landing page and/or were logged in to their CoLab account. That way, if someone was sent a Social Network Prize link and then became a member, we could associate the referral chain to the individual's membership. Only if the person was sent the referral link and later remembered the CoLab and Googled it on a different browser (i.e., instead of using the link) would they *not* be associated with their referral chain. Therefore determining whether a CoLab member who engaged with the Social Network Prize was an existing member or a new member recruited by the Social Network Prize entailed a simple comparison between the time the person first clicked on a Social Network Prize link and the time that they first created their membership.

While the Social Network Prize was re-launched in 2015 and 2016 and we include that data in our analysis, the majority of data for this case study comes from 2014 when staff made the most effort to spread the word of the contest.

## 3   DATA

### 3.1   The Referral Network

Over 36,000 people clicked on a Social Network Prize link and over 1,000 of those people entered their email address to generate their own unique link, which they could then share with their friends (a visualization of the spread of the Social Network Prize is depicted in Figure 4). Of the 36,000+ people who clicked on a Social Network Prize link, 351 people who were not already Climate CoLab members decided to join. The vast majority of these newly recruited members (309 of 351) were recruited by someone already directly affiliated with Climate CoLab—either by a current member or by a staff-generated link, but there were 42 new members who were indirectly recruited (i.e., by someone who was not a member when they clicked on the Social Network Prize link themselves). In other words, someone who was also recruited by the Social Network Prize recruited these specific 42 new members.

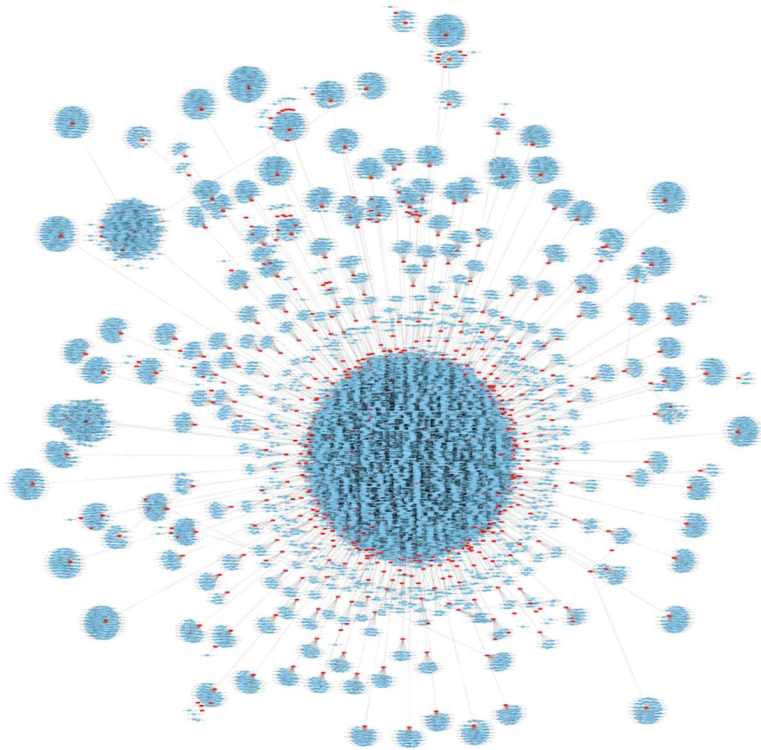

Figure 4: Visualization of all people who engaged with the Social Network Prize contest. Blue points indicate people who clicked on a Social Network Prize link. Red points indicate people who, upon clicking on a Social Network Prize link, entered their email address to create their own unique link that they could share with others. Lines connect a person to the person who shared the link with them, and the cluster in the center represents people who were one degree of separation from a CoLab member (i.e., people who were directly referred). The recursive incentive mechanism of the Social Network Prize provided an incentive for some people with whom the link was shared to share it with others as well, allowing it to spread further into other social networks.

### 3.2 Recruited Members Proposal Activity

The 351 newly recruited members made up a small portion (less than 1%) of the entire CoLab community of 78,390 members. However, as shown in Table 1, they contributed an outsized number of proposals: 57 of the 1284 proposal authors were newly recruited members from the Social Network Prize contests. Moreover, many of the proposals from these newly recruited members were high quality proposals that became finalists or even won their respective contests.

Table 1

|  | Users | Proposal Authors | Finalists | Winners |
|---|---|---|---|---|
| **Social Network Prize New Member Recruits** | 351 | 57 | 16 | 9 |
| → Directly Recruited (i.e., referred by current member or staff-generated link) | 309 | 52 | 13 | 7 |
| → Indirectly Recruited (i.e., referred by someone who was referred by someone) | 42 | 5 | 3 | 2 |
| **Other CoLab Members** | 78039 | 1227 | 228 | 120 |
| **TOTAL CoLab Members** | 78390 | 1284 | 244 | 129 |

### 3.3 Participant Characteristics

For people who engaged with the Social Network Prize but did not become CoLab members, the only information we have is their IP address. From this information, we know that this group is highly geographically diverse: the 36,000+ people who clicked on a Social Network Prize Link were from over 100 different countries, the 1,000+ people who entered their email address were from over 60 different countries, and the 351 newly recruited members were from over 40 different countries.

We have more information on the people who were actually recruited through the Social Network Prize based on surveys sent to all CoLab members. The 2,607 unique survey responses from the entire CoLab community indicate that the community is male skewed (60%), has a median age of 30-39, is highly educated (55% attend or completed graduate school), and from all over the world (54% from outside of the United States). We have survey responses from 55 members who were recruited through the Social Network Prize, and these members were similar to the rest of the community in terms of gender (69% male), median age (30-39), and education (53% attend or completed graduate school). However, 80% of the newly recruited members live outside of the United States, which is significantly more than other survey respondents, $\chi^2(1) = 15.50$, $p < .001$.

While we do not have enough survey data on Social Network Prize recruits to make many meaningful comparisons between the characteristics of indirect vs. direct recruits, or finalists to non-finalists, it is notable that 8 finalists who were recruited by the Social Network Prize responded to the survey and all of them live outside of the United States.

### 4 RESULTS

### 4.1 Social Network Prize Recruits vs. Other CoLab Members

We first set out to compare the activity of CoLab members recruited via the Social Network Prize to that of other members. Whereas only 1.6% (1,227 of 78,039) of other members submitted proposals to CoLab contests, over 16% (57 of 351) of members recruited by the Social Network Prize did, which is a highly significant difference, $\chi2(1) = 137353$, $p < .001$.

We next analyzed whether members recruited via the Social Network Prize were more likely to become finalists. While 4.6% of members recruited by the Social Network Prize became finalists, only .003% of other members did, a highly significant difference, $\chi^2(1) = 130757$, $p < .001$.

We also analyzed whether members recruited via the Social Network Prize were more likely to become finalists conditional on having submitted a proposal. While 28.1% of authors recruited by the Social Network Prize (16 of 57) became finalists, only 18.6% of other authors (228 of 1227) became finalists. This difference is marginally statistically significant, $\chi2(1) = 3.19$, $p = .074$.

### 4.2 Direct vs. Indirect Social Network Prize Recruits

We next set out to compare the activity of newly recruited CoLab members who were recruited directly by a member or a staff-generated link with those who were recruited indirectly (i.e., by someone who was also recruited via the Social Network Prize). While 16.8% (52 of 309) of directly-recruited members authored proposals, only 11.9% (5 of 42) of indirectly-recruited members did so. On account of the small sample sizes, we employed a Fisher's Exact Test instead of a chi-square and found that this difference was not statistically significant, $p = .509$.

Finally, we analyzed whether directly-recruited members were more likely to become finalists, conditional on having submitted a proposal, than indirectly-recruited members. While 25% of directly recruited authors became finalists (13 of 52), 60% of indirectly recruited proposal authors became finalists (3 of 5). Again employing a Fisher's Exact Test on account of the small sample size, we obtain a test statistic value of $p = .129$, providing modest evidence that indirect recruits author higher quality proposals than indirect recruits. Notably, two of the three finalist proposals from an indirect recruit won their respective contests (compared to 7 of 13 from direct recruits) and one of these was awarded the $10,000 Grand Prize in 2014, triggering financial awards for the Social Network Prize.

## 5 DISCUSSION

In our Social Network Prize case study, we find evidence that new members who were brought in by referrals were more likely to author proposals—and to author finalist proposals—than other CoLab members. We also find suggestive evidence that indirect recruits were more likely to author high quality proposals than direct recruits, which is consistent with the idea that innovative solutions to complex problems are likely to come from people in distant networks who bring in new information and perspectives. For instance, it is possible that by the time the Social Network Prize was launched, many pre-existing members had already told many of their close friends and colleagues with the best ideas for addressing climate change to join the CoLab, thereby exhausting the innovative potential of their local networks. However, the recursive incentive structure of the Social Network Prize opened up a new group of people who could be referred: people who might not have ideas themselves, but who might be likely to know other people with ideas. By incentivizing these weak ties, the Social Network Prize may have been able to mobilize not only more people than a traditional referral incentive, but also more people with new information and perspectives. This theoretical account is also supported by the survey data analysis, which indicates that users recruited from the Social Network Prize were significantly more likely to live outside the United States than other members.

Of course, our work has some limitations. Importantly, we only consider one particular case of leveraging a recursive incentive mechanism for promoting an open-innovation contest, and we do not know how well this result will generalize to other settings. We also have a relatively small sample size in terms of the number of new recruits, and especially the number of indirect recruits. Furthermore, we do not know enough about the characteristics of the recruits who authored finalist proposals to draw meaningful conclusions about them, although it is noteworthy that all eight for whom we have survey data live outside the United States. Therefore, while we know from other research that 1) recursive incentive mechanisms are a useful tool for accessing people with novel information and alternative viewpoints [18], and 2) such people often submit innovative ideas to open innovation contests [6-7], we are unable to prove that this is why the proposals of Social Network Prize recruits were of high quality.

However, there are also reasons why our results may underestimate the potential effectiveness of recursive incentive schemes for social network recruitment. First, it can be difficult to quickly communicate how the recursive incentive scheme works because it is so much less common than traditional referral programs. But if recursive incentive schemes for social network recruitment become more common, this barrier to implementation will become less significant. Second, a desirable feature of the recursive incentive scheme is that it only costs an implementer more money than a traditional referral program *if it works*. This is precisely because of the recursive incentive structure. In other words, intermediaries are only compensated if they are critical for finding the end target. For these reasons, we believe that our findings are compelling enough that a wide range of organizations—especially those pursuing open innovation initiatives—might want to consider experimenting with referral programs that include recursive incentives.